\documentclass[12pt, a4paper]{article}
\usepackage{amsmath}
\everymath{\displaystyle}
\usepackage{mathtools}
\usepackage{amsfonts}
\usepackage{amssymb}
\usepackage{amsthm}
\usepackage[utf8]{inputenc}
\usepackage[toc]{appendix}
\usepackage[hiresbb]{graphicx}
\usepackage{caption}
\usepackage{subcaption}
\usepackage{longtable}
\usepackage{booktabs}
\usepackage{listings}
\usepackage{here}
\usepackage{float}
\usepackage{ascmac}
\usepackage{tikz}
\usepackage{url}
\usepackage{lipsum}
\usepackage{authblk}
\usepackage{eqnarray}
\usepackage{siunitx}

\usepackage[sectionbib,round]{natbib}

\newcommand{\be}{\begin{equation}}
\newcommand{\ee}{\end{equation}}
\newcommand{\beq}{\begin{eqnarray*}}
\newcommand{\eeq}{\end{eqnarray*}}
\def\sym#1{\ifmmode^{#1}\else\(^{#1}\)\fi}

\title{\Large{\bf{Wavelet Analysis of Cryptocurrencies 
--- Non-Linear Dynamics in High Frequency Domains}}}
\author{\large{\bf{Tatsuru Kikuchi}}}
\affil{\small{\it{Faculty of Economics, The University of Tokyo,}}\\
{\it{7-3-1 Hongo, Bunkyo-ku, Tokyo 113-0033 Japan}}}
\date{\small{(\today)}}
\begin{document}
\maketitle
\begin{abstract}
In this study, we perform some analysis for the probability distributions in the space of frequency and time variables. However, in the domain of high frequencies, it behaves in such a way as the highly non-linear dynamics. The wavelet analysis is a powerful tool to perform such analysis in order to search for the characteristics of frequency variations over time for the prices of major cryptocurrencies. In fact, the wavelet analysis is found to be quite useful as it examine the validity of the efficient market hypothesis in the weak form, especially for the presence of the cyclical persistence at different frequencies. If we could find some cyclical persistence at different frequencies, that means that there exist some intrinsic causal relationship for some given investment horizons defined by some chosen sampling scales. This is one of the characteristic results of the wavelet analysis in the time-frequency domains. 
\end{abstract}
\newpage
\section{Introduction}
In our previous study in \citet{TK2021}, we found a positive evidence of price stability of cryptocurrencies as a medium of exchange. In fact, for the sample years from 2016 to 2020, the prices of major cryptocurrencies have been found to be stable, relative to major financial assets. In our previous study, the main issues what we have accomplished the study is in understanding of the overall behavior or trend of the prices of cryptocurrencies. In other words, it is necessary to remove noisy, high-frequency behaviors by introducing high-frequency filters. 

However, when we seek for the characteristics which differs in different cryptocurrencies, we found that high-frequency variances can become one of the indicators that shows the specific behaviors for each cryptocurrencies. In this study, we perform some analysis for the probability distributions in the space of frequency and time variables. However, in the domain of high frequencies, it behaves in such a way as the highly non-linear dynamics. The wavelet analysis is a powerful tool to perform such analysis in order to search for the characteristics of frequency variations over time for the prices of major cryptocurrencies. Using a wavelet transform, the wavelet compression methods are adequate for representing transients, such as high-frequency components in two-dimensional images. 

It is known that in the Hamiltonian formulation of classical mechanics or sympletic geometry in differential geometry, the set of all possible configurations of a system is modeled as a sympletic manifold, and this manifold's cotangent bundle describes the phase space of the system. Along the lines of the sympletic geometry, the dual space of the vector space which represents the 'time' variable is nothing but the energy or equivalently the frequency space, that is, the phase space of the system in concern is spanned by the time and frequency variables. 

There have already been many studies by the use of wavelet analysis in the literature of cryptocurrencies. \citet{Bo2020} have applied the wavelet coherency analysis and showed the overall dependence between Bitcoin/gold/commodities and the stock markets is not very strong at various time scales, with Bitcoin being the least dependent. \citet{Fru2021} have investigated on the bubble behavior of bitcoin (BTC) whose price has a peak in 2017. They have performed a structural change point analysis in order to identify the peak of BTC price in 2017. \citet{Li2021} have made an investigation on the relationship between investor attention and the cryptocurrency markets. They gave proposed the combination of Granger causality test and wavelet analysis to analyze the causal relation between investor attention and cryptocurrencies. \citet{Go2021} have accomplished a wavelet analysis to daily data of COVID-19 world deaths and daily Bitcoin prices from 13th December 2019 to 29th April 2020. They found, especially for the period post April 5, that levels of COVID-19 caused a rise in Bitcoin prices. \citet{Ar2022} have examined a study of the long term memory in return and volatility, using high frequency time series of seven major cryptocurrencies. They proposed a new wavelet analysis which provides more robust estimators of the Hurst exponent. It is examined their analysis during the period of Covid-19. They found that, during the peak of Covid-19 pandemic, the long memory of returns was only mildly affected. However, volatility suffered a temporary impact in its long range correlation structure. \citet{Pa2024} have proposed a way to make price predictions of cryptocurrencies. They have used wavelet transform to split a non-stationary signal into uncorrelated components by extracting the features from an available data. 

Not only in the context of digital finance, the wavelet analysis has also been used in economics. \citet{Gen2001} have emphasized the ability of the wavelet analysis, specifically in the estimation of volatility models since it can be used to extract unnecessary, seasonality components in the market data. They proposed a method for intraday seasonality extraction that is free of model selection parameters which may affect other intraday seasonality filtering methods. Their methodology is based on a wavelet multi-scaling approach which decomposes the data into its low- and high-frequency components through the application of a non-decimated discrete wavelet transform. \citet{Ag2008} have used wavelet analysis to analyze the impact of interest rate price changes on some macroeconomic variables: industrial production, inflation and the monetary aggregates M1 and M2. They emphasized that one of the advantages of the wavelet analysis is the possibility of uncovering transient relations. Moreover, the same conclusion is reached about inflation and the volatility of inflation decreased at the same time. \citet{Ru2009} have performed a wavelet analysis to measure the co-movement among international stock markets. They found a way to characterize how international stock returns relate in the time and frequency domains simultaneously, which allows one to provide a richer analysis of the co-movement. \citet{Re2013} have made an investigation on the relationship between oil prices and US dollar exchange rates using wavelet multi-resolution analysis. They characterized the oil price–exchange rate relationship for different timescales in an attempt to disentangle the possible existence of contagion and interdependence during the global financial crisis and analyze possible lead and lag effects.

\section{Data and Basic Methodology}
As similar to our previous studies, we take three major cryptocurrencies, that is, Bitcoin (BTC), Ethereum (ETH) and Ripple (XRP). We extract the historical daily exchange prices of those cryptocurrencies through the API (Application Programming Interface) provided by the CryproCompare, who is one of the leading global cryptocurrency market data providers. Just because of the limitation of the API service, whose maximal length of the data we can extract at a time is limited to be at most 2000. Taking such a limitation into account, the study examines the daily closing prices for three major cryptocurrencies (BTC, ETH and XRP) from July 10, 2017 to December 31, 2022 through the API of the CryproCompare. The original closing prices $p_{t}$ at time $t$ is converted to the log returns as follows. 
\be
x(t) = \log \left( \frac{p_{t}}{p_{t-1}}  \right) \;,
\ee
where $p_{t}$ and $p_{t-1}$ are the current and previous closing prices, respectively. 

Moreover, we use the historical daily exchange data for S\&P500 index and GOLD price, as well as the historical daily exchange data for the JPY/USD (Japanese Yen and US Dollar) and the USD/EUR (US Dollar and Euro) currency pairs with the period of time from July 10, 2017 to December 31, 2022. The data for S\&P500 index is extracted from the FRED (Federal Reserve Economic Data) and the data for GOLD/USD, JPY/USD and USD/EUR are taken from the web site, 'investing.com' (https://www.investing.com/). 

The wavelet analysis is quite useful as it examine the validity of the efficient market hypothesis in the weak form, especially for the presence of the cyclical persistence at different frequencies. If we could find some cyclical persistence at different frequencies, that means that there exists some intrinsic causal relationship for some given investment horizons defined by some chosen sampling scales. This is one of the characteristic results of the wavelet analysis in the time-frequency domains. 

The basic methodology of our analysis is nothing but the standard wavelet analysis. However, for the sake of defining our notations, we give a brief summary of wavelet transformation. Let us consider the Hilbert space ${\cal{H}} = L^{2}({\mathbb{R}})$, a set of square integrable functions. For a given function $f \in {\cal{H}} = L^{2}({\mathbb{R}})$, the continuous wavelet transform is defined by
\be
W_{f}(\sigma, \tau) = \frac{1}{|\sigma|^{1/2}} \int_{-\infty}^{\infty} \bar{\psi} \left(\frac{t-\tau}{\sigma} \right) f(t) dt \;,
\ee
where $\sigma \in \mathbb{R}$ is called the scaling parameter, and $\tau \in \mathbb{R}$ is the translation parameter, and $\psi \in L^{2}({\mathbb{C}})$ is a continuous function, which is sometimes called the mother wavelet or mother function. The wavelet transform can provide us with the frequency of the signals and the time associated to those frequencies, making it very convenient for its application in numerous fields. 

In this study, we mainly use the real Morlet wavelet (or Morlet wavelet) and the complex Morlet (or Gabor wavelet) as the mother functions, both of which are composed of a complex exponential multiplied by a Gaussian window. In case of the Morlet wavelet, the mother wavelet is given by
\be
\psi(t) = \pi^{-1/4} e^{-t^2/2} e^{i \omega t} \;,
\ee
where $\omega$ is a fixed frequency parameter, which is normally taken to be 6 to satisfy the admissibility condition. For simplicity, we normally denote the Morlet wavelet transform as 'morl' in the outputs. 

On the other hand, in case of the complex Morlet wavelet, the mother function is defined by
\be
\psi(t) = (\pi \, \delta )^{-1/4} e^{-t^2/\delta} e^{i \omega t} \;,
\ee
where $\omega$ is a fixed frequency parameter and $\delta$ controls the decay in the time domain and the corresponding energy spread (bandwidth) in the frequency domain. For simplicity, we normally denote the complex Morlet wavelet transform as 'cmor$\delta$-$\omega$' with some given floating values $\delta$ and $\omega$, up to the normalization. Essentially because $\delta$ is the inverse of the variance in the frequency domain, increasing the value of $\delta$ makes the wavelet energy more concentrated around the center frequency and results in slower decay of the wavelet in the time domain and decreasing the value of $\delta$ results in faster decay of the wavelet in the time domain and less energy spread in the frequency domain. 

As similar to the Fourier analysis, we can define some useful measures, one of which is a cross-wavelet spectrum which captures the covariance between two time series in the time and frequency space. For given two time series $x(t)$ and $y(t)$ with those wavelet transforms $W_{x}(\sigma, \tau)$ and $W_{y}(\sigma, \tau)$, the cross-wavelet is defined by $W_{x y}(\sigma, \tau) = W_{x}(\sigma, \tau) W_{y}^{*}(\sigma, \tau)$. Based on the cross-wavelet, we can define another useful measure, the wavelet coherence which is defined by 
\be
R^{2}(\sigma, \tau) = \frac{\Big| W_{x y}(\sigma, \tau) \Big|^{2}}{ \Big| W_{x}(\sigma, \tau) \Big|^{2} \Big| W_{y}(\sigma, \tau) \Big|^{2} } \;.
\ee

\section{Result}
\begin{figure}[H]
	\includegraphics[width=1.0\textwidth]{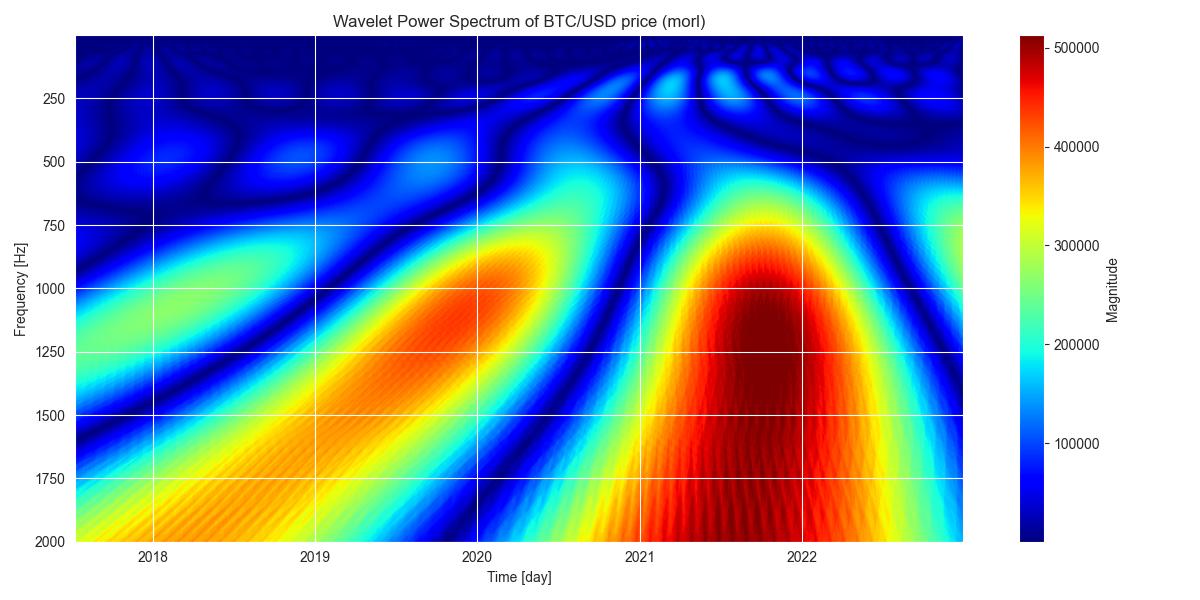}
	\caption{Wavelet Power Spectrum of BTC/USD (morl)}
	\label{fig:Fig1}
\end{figure}

\begin{figure}[H]
	\includegraphics[width=1.0\textwidth]{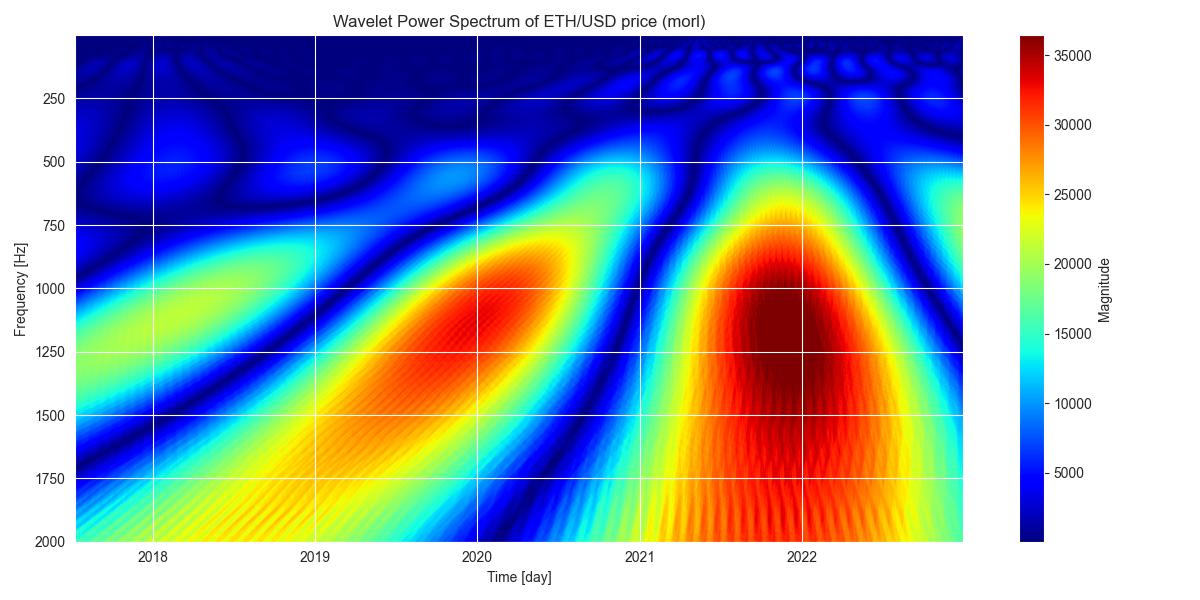}
	\caption{Wavelet Power Spectrum of ETH/USD (morl)}
	\label{fig:Fig2}
\end{figure}

\begin{figure}[H]
	\includegraphics[width=1.0\textwidth]{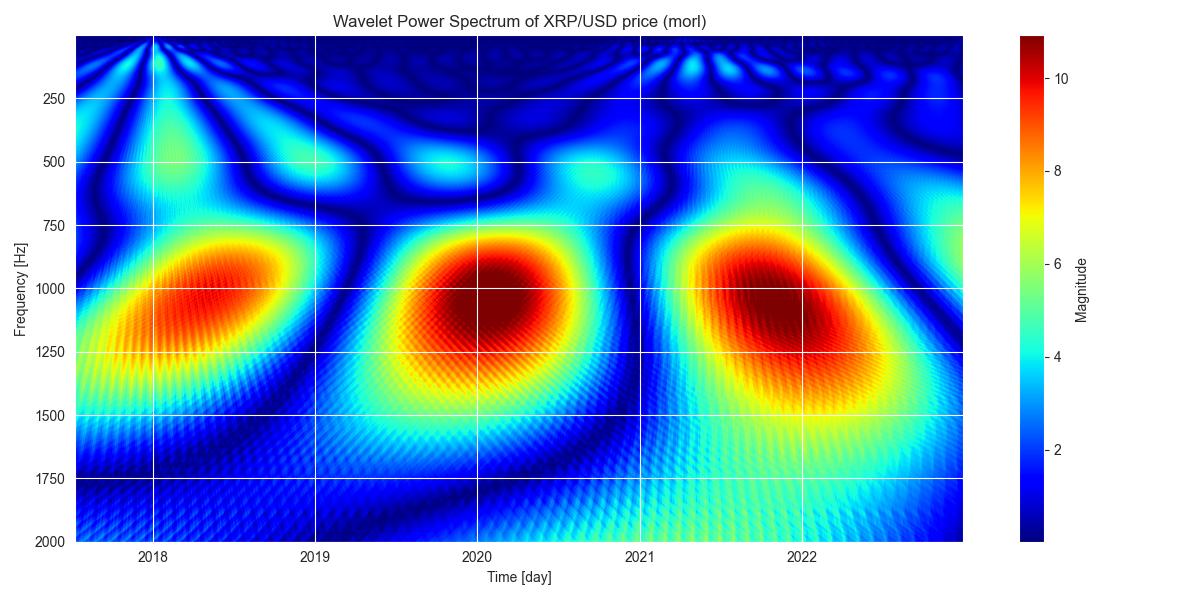}
	\caption{Wavelet Power Spectrum of XRP/USD (morl)}
	\label{fig:Fig3}
\end{figure}

\begin{figure}[H]
	\includegraphics[width=1.0\textwidth]{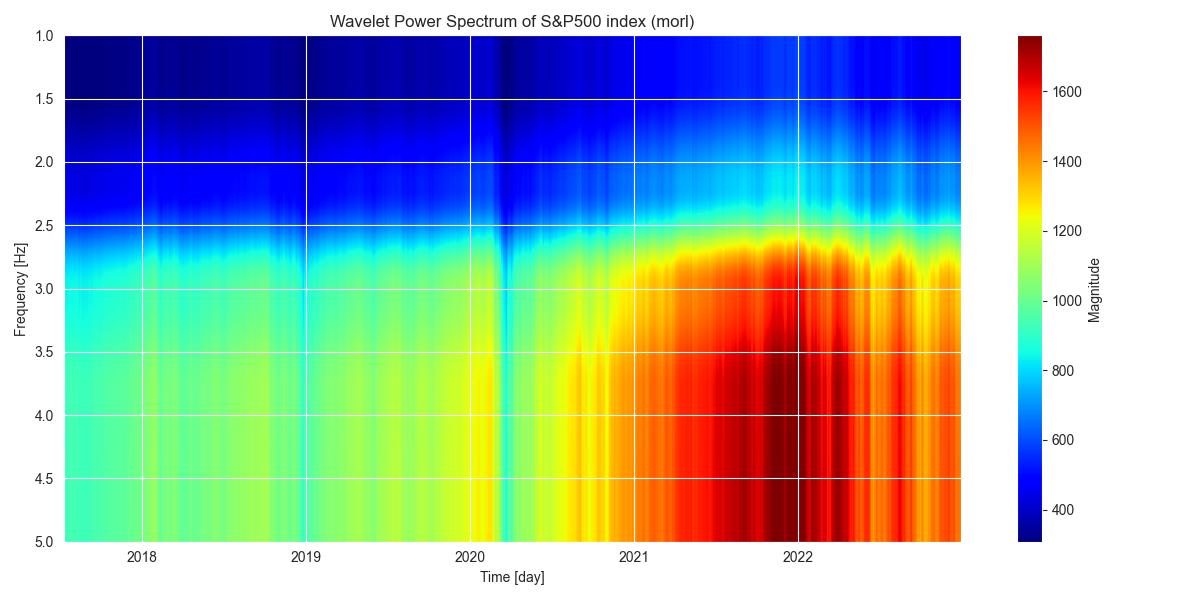}
	\caption{Wavelet Power Spectrum of S\&P500 index (morl)}
	\label{fig:Fig4}
\end{figure}

\begin{figure}[H]
	\includegraphics[width=1.0\textwidth]{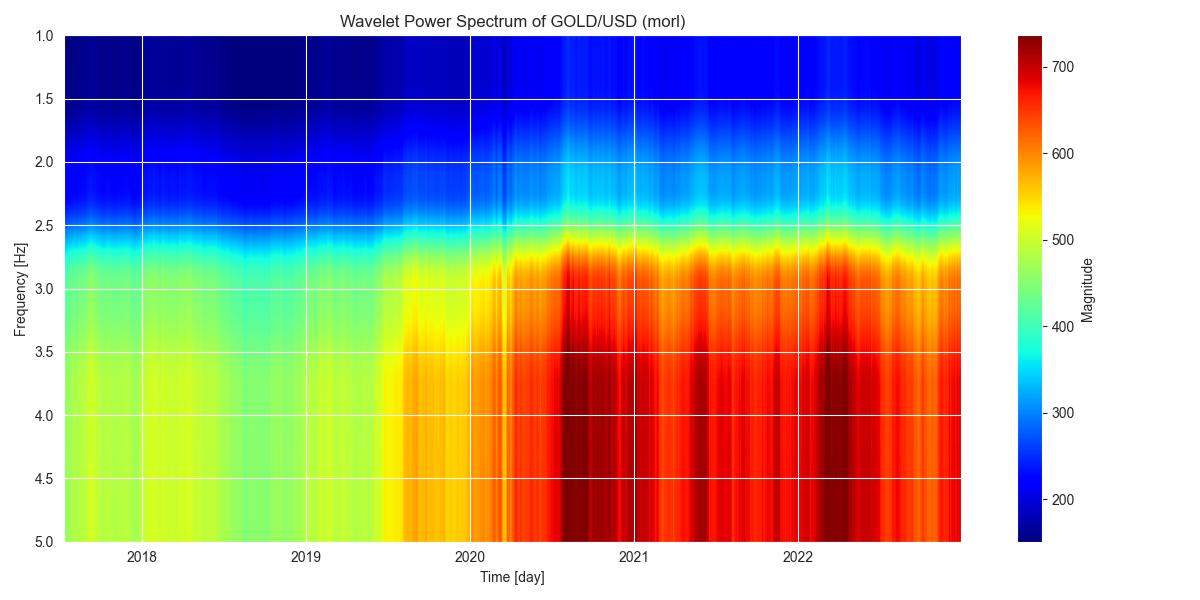}
	\caption{Wavelet Power Spectrum of GOLD/USD (morl)}
	\label{fig:Fig5}
\end{figure}

\begin{figure}[H]
	\includegraphics[width=1.0\textwidth]{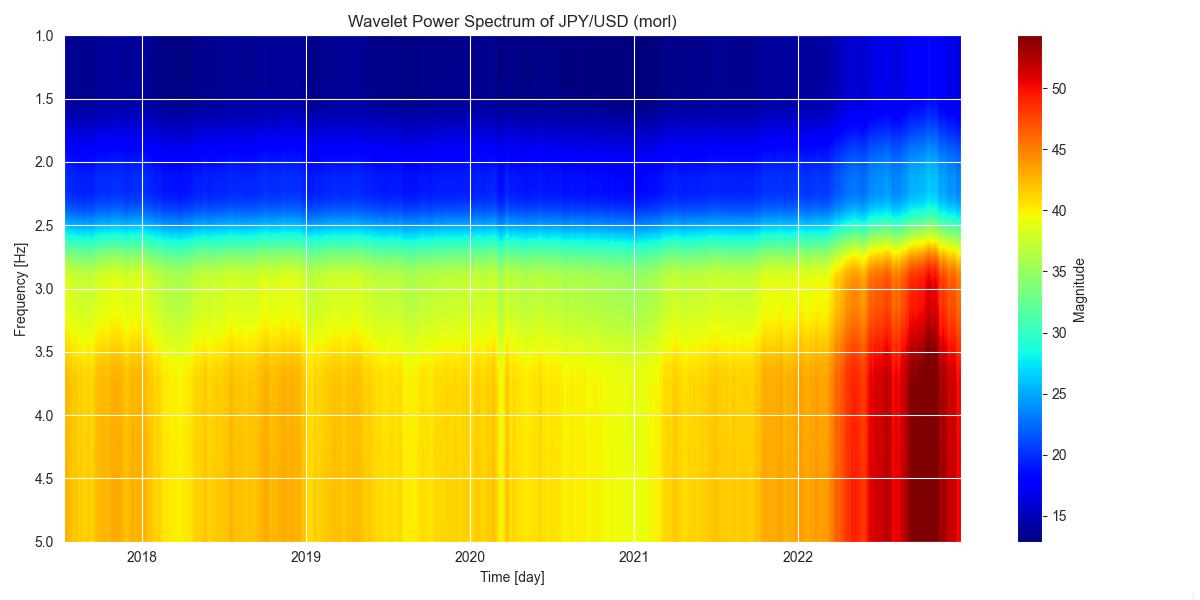}
	\caption{Wavelet Power Spectrum of JPY/USD (morl)}
	\label{fig:Fig6}
\end{figure}

\begin{figure}[H]
	\includegraphics[width=1.0\textwidth]{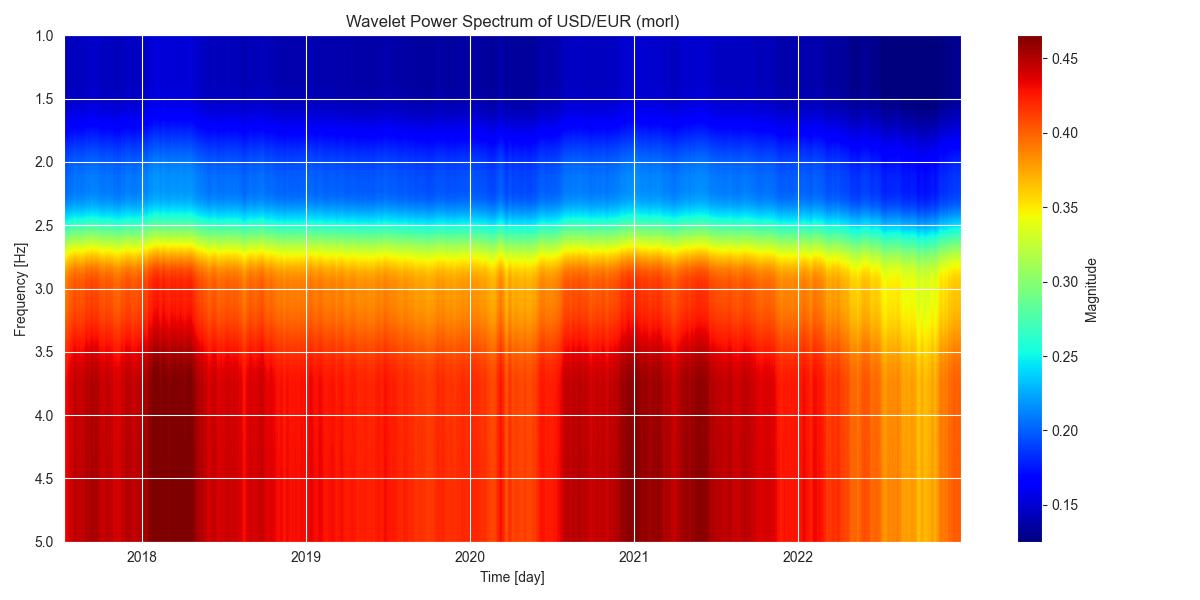}
	\caption{Wavelet Power Spectrum of USD/EUR (morl)}
	\label{fig:Fig7}
\end{figure}

The results of our wavelet analysis based on the Morlet wavelet transform for the cryptocurrencies (BTC, ETH, XRP), S\&P500 index, GOLD/USD, JPY/USD and USD/EUR prices are shown in Figure \ref{fig:Fig1}, Figure \ref{fig:Fig2}, Figure \ref{fig:Fig3}, Figure \ref{fig:Fig4}, Figure \ref{fig:Fig5}, Figure \ref{fig:Fig6} and Figure \ref{fig:Fig7}. It shows that if we focus on the low-frequency region, it is stable enough as the time changes for any cryptocurrencies we are interested in. However, if we take the high-frequency region, that is in the bottom line of each figures, we can see the 'hot spots' as represented by the red color. Those hot spots corresponds to the peaks in the standard, one-dimensional time series analysis. One of the most intriguing observation is that during the period of Covid-19 peek in 2022, all the cryptocurrencies show the hot spots in the wavelet analysis. Second curious observation is that there are sequential points, or almost like a line in the case of XRP price, in the low to middle range of frequency regions as the time changes. Such a line-like shape shows the existence of some unique frequency or periodicity for the price of cryptocurrencies. 

\begin{figure}[H]
	\includegraphics[width=1.0\textwidth]{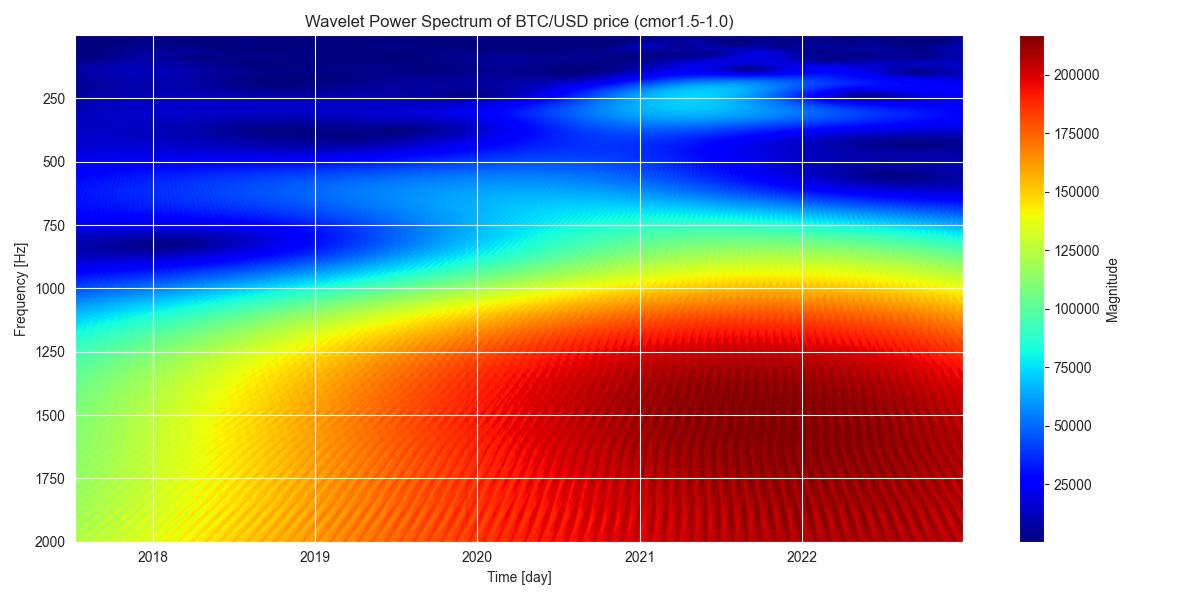}
	\caption{Wavelet Power Spectrum of BTC/USD (cmor1.5-1.0)}
	\label{fig:Fig8}
\end{figure}

\begin{figure}[H]
	\includegraphics[width=1.0\textwidth]{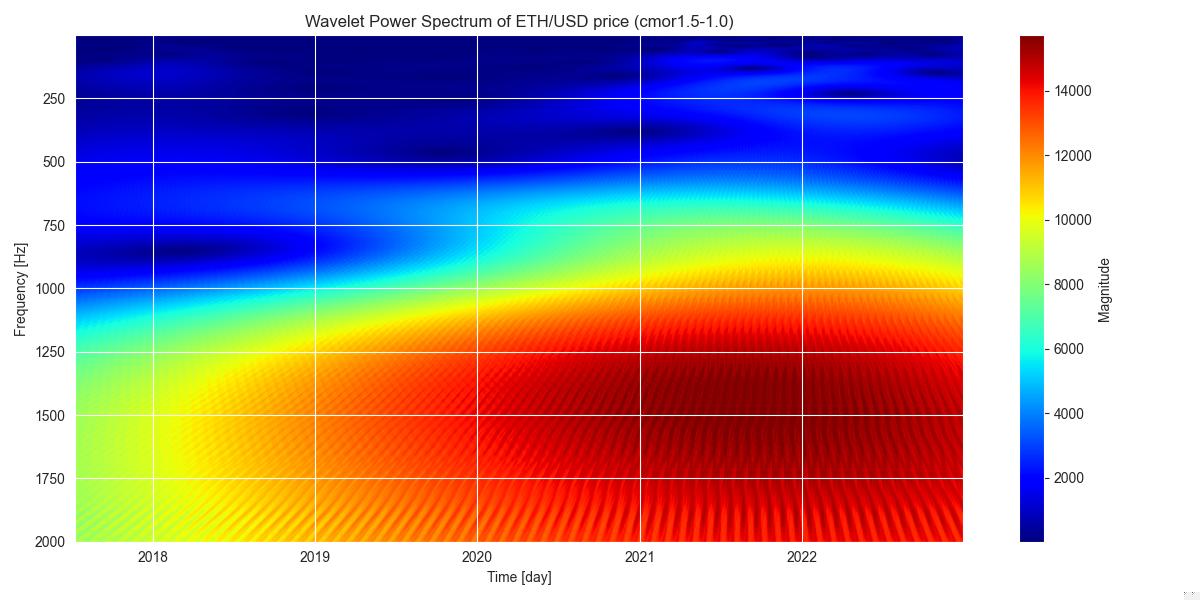}
	\caption{Wavelet Power Spectrum of ETH/USD (cmor1.5-1.0)}
	\label{fig:Fig9}
\end{figure}

\begin{figure}[H]
	\includegraphics[width=1.0\textwidth]{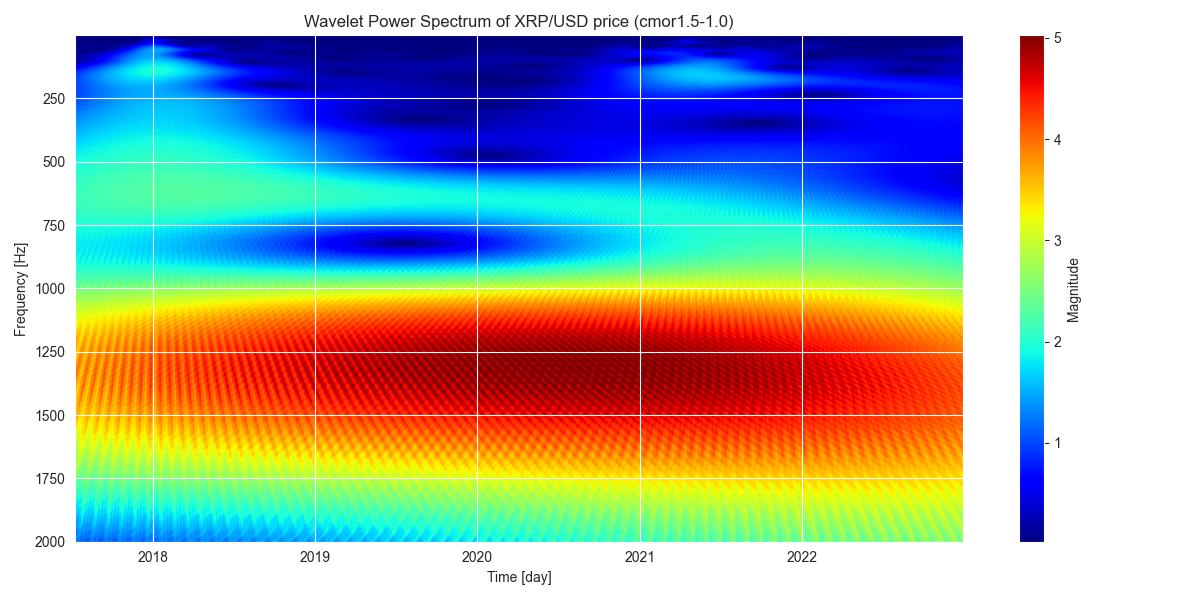}
	\caption{Wavelet Power Spectrum of XRP/USD (cmor1.5-1.0)}
	\label{fig:Fig10}
\end{figure}

\begin{figure}[H]
	\includegraphics[width=1.0\textwidth]{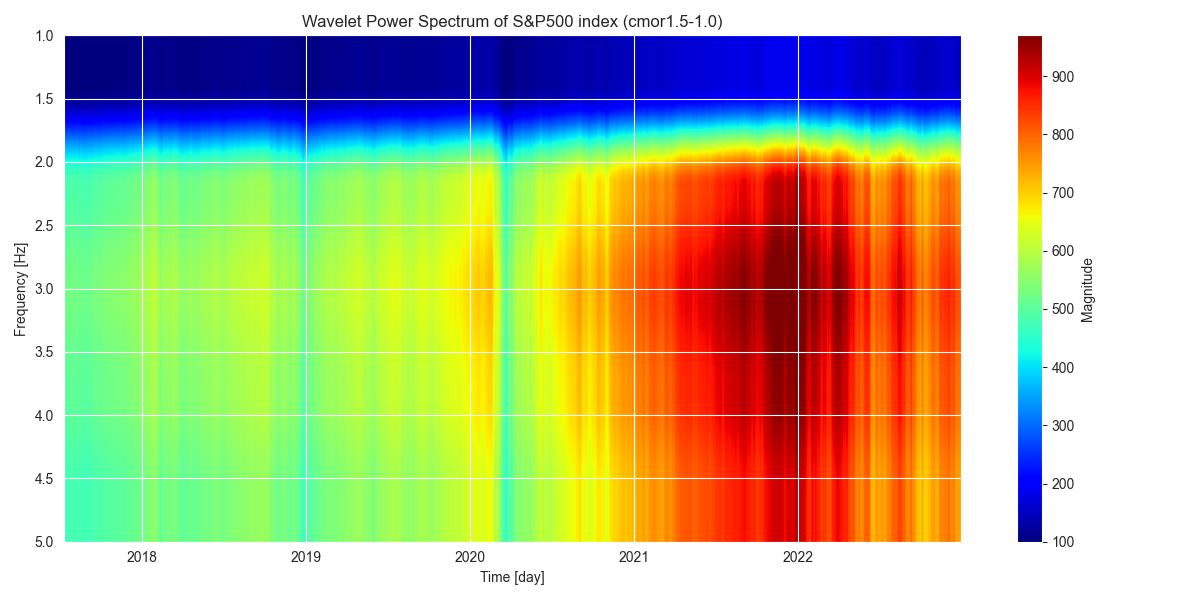}
	\caption{Wavelet Power Spectrum of S\&P500 index (cmor1.5-1.0)}
	\label{fig:Fig11}
\end{figure}

\begin{figure}[H]
	\includegraphics[width=1.0\textwidth]{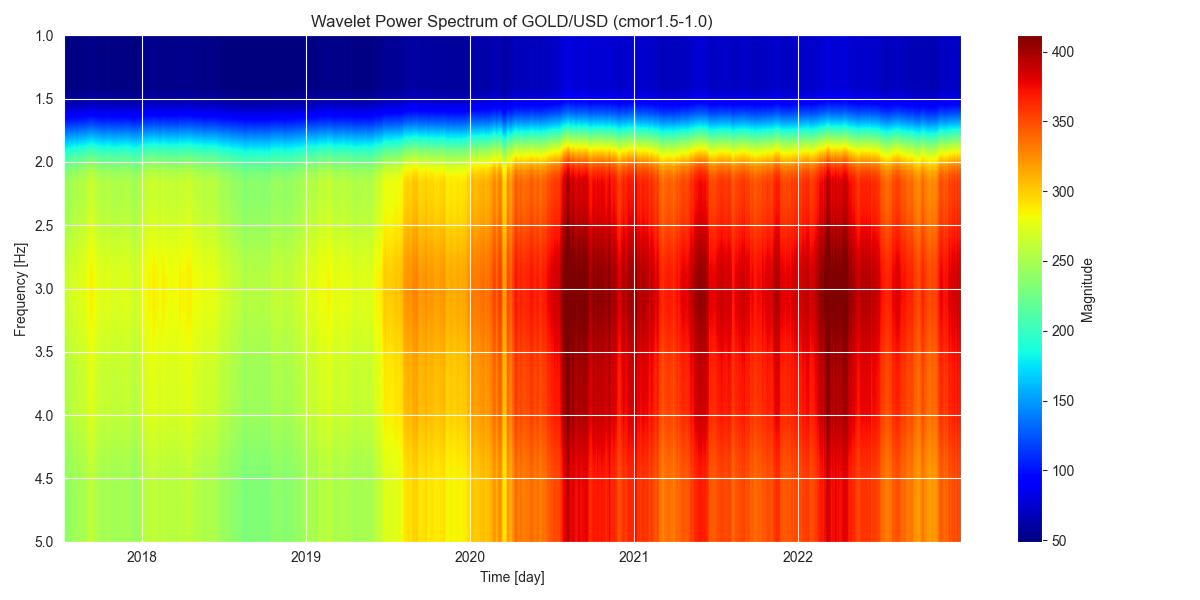}
	\caption{Wavelet Power Spectrum of GOLD/USD (cmor1.5-1.0)}
	\label{fig:Fig12}
\end{figure}

\begin{figure}[H]
	\includegraphics[width=1.0\textwidth]{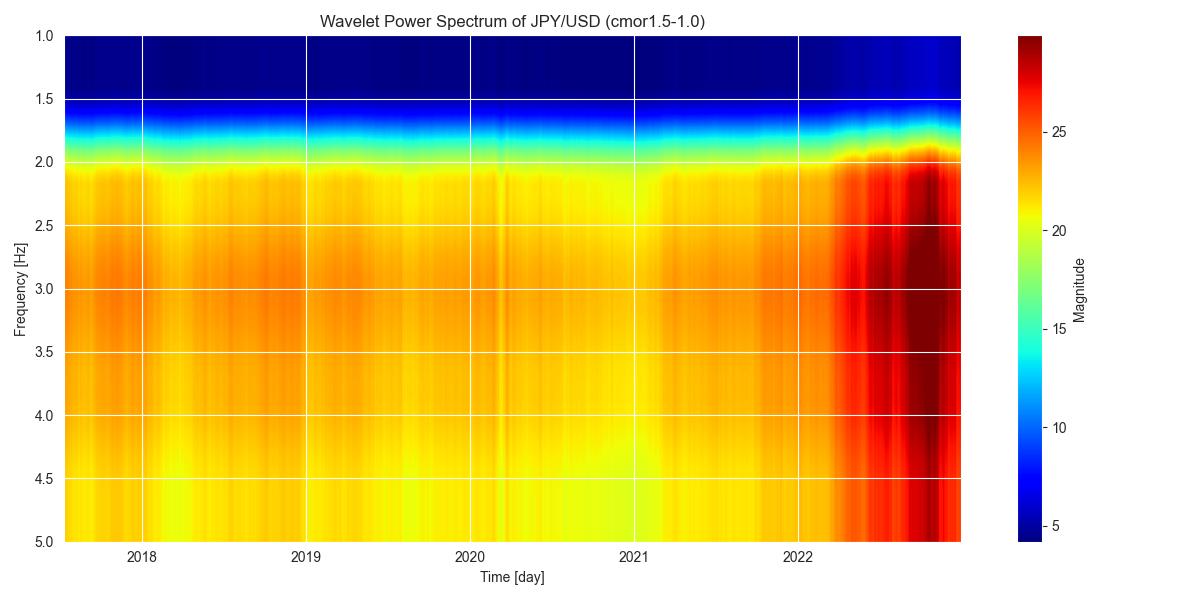}
	\caption{Wavelet Power Spectrum of JPY/USD (cmor1.5-1.0)}
	\label{fig:Fig13}
\end{figure}

\begin{figure}[H]
	\includegraphics[width=1.0\textwidth]{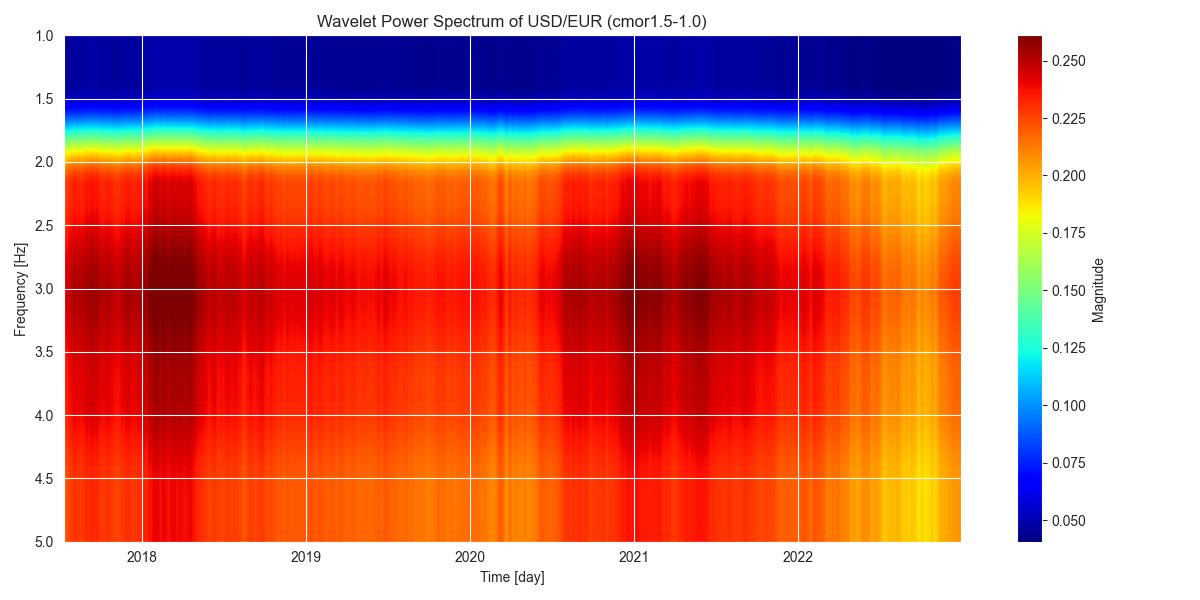}
	\caption{Wavelet Power Spectrum of USD/EUR (cmor1.5-1.0)}
	\label{fig:Fig14}
\end{figure}

On the other hand, for the cryptocurrencies (BTC, ETH, XRP), S\&P500 index, GOLD/USD, JPY/USD and USD/EUR prices, the results of our wavelet analysis based on the complex Morlet wavelet transform are shown in Figure \ref{fig:Fig8}, Figure \ref{fig:Fig9}, Figure \ref{fig:Fig10}, Figure \ref{fig:Fig11}, Figure \ref{fig:Fig12}, Figure \ref{fig:Fig13} and Figure \ref{fig:Fig14}. It also shows that if we focus on the low-frequency region, it is stable enough as the time changes for any cryptocurrencies we are interested in. However, if we take the high-frequency region, that is in the bottom line of each figures, we can see the 'hot spots' as represented by the red color. 

\section{Conclusion}
We have performed the wavelet analysis, which is well-established powerful tool to perform an analysis in order to search for the characteristics of frequency variations over time, to make investigations in the time-frequency domains for the prices of major cryptocurrencies and major stock market indicators such as S\&P500, Gold, exchange data of JPY/USD and USD/EUR. In fact, the wavelet analysis is found to be quite useful as it examine the validity of the efficient market hypothesis in the weak form, especially for the presence of the cyclical persistence at different frequencies. We found some cyclical persistence at different frequencies, especially for the daily exchange data of XRP cryptocurrency. It means that there exists some intrinsic causal relationship for some given investment horizons defined by some chosen sampling scales. This is one of the characteristic results of the wavelet analysis in the time-frequency domains. 

The result of our wavelet analysis shows that if we focus on the low-frequency region, it is stable enough as the time changes for any cryptocurrencies we are interested in. However, if we take the high-frequency region, that is in the bottom line of each figures, we can see the 'hot spots' as represented by the red color. Those hot spots corresponds to the peaks in the standard, one-dimensional time series analysis. One of the most intriguing observation is that during the period of Covid-19 peek in 2022, all the cryptocurrencies show the hot spots in the wavelet analysis. Second curious observation is that there are sequential points, or almost like a line in the case of XRP price, in the low to middle range of frequency regions as the time changes. Such a line-like shape shows the existence of some unique frequency or periodicity for the price of cryptocurrencies.

\end{document}